\documentclass[twocolumn]{aastex631}
\usepackage{amsmath, amsthm, amssymb}

\usepackage{multirow}
\shorttitle{Transiting Exosatellites with Spitzer}
\shortauthors{Limbach et al.}
\graphicspath{{./}{figures/}}

\begin{document}
\title{Occurrence Rates of Exosatellites Orbiting 3--30\,M$_{\rm Jup}$ Hosts from 44 Spitzer Light Curves}

\correspondingauthor{Mary Anne Limbach}
\email{maryannelimbach@gmail.com}

\author[0000-0002-9521-9798]{Mary Anne Limbach}
\affiliation{Department of Astronomy, University of Michigan, Ann Arbor, MI 48109, USA}

\author[0000-0003-0489-1528]{Johanna M. Vos}
\affiliation{School of Physics, Trinity College Dublin, The University of Dublin, Dublin 2, Ireland}
\affiliation{Department of Astrophysics, American Museum of Natural History, Central Park West at 79th Street, NY 10024, USA}

\author[0000-0001-7246-5438]{Andrew Vanderburg}
\affiliation{Department of Physics and Kavli Institute for Astrophysics and Space Research, Massachusetts Institute of Technology, Cambridge, MA 02139, USA}

\author[0000-0002-8958-0683]{Fei Dai} 
\affiliation{Division of Geological and Planetary Sciences, California Institute of Technology,
1200 E California Blvd, Pasadena, CA, 91125, USA}

\begin{abstract}
We conduct a comprehensive search for transiting exomoons and exosatellites within 44 archival Spitzer light curves of 32 substellar worlds with estimated masses ranging between 3--30\,M$_{\rm Jup}$. This sample's median host mass is 16\,M$_{\rm Jup}$, inclusive of 14 planetary-mass objects, among which one is a wide-orbit exoplanet. We search the light curves for exosatellite signatures and implement a transit injection-recovery test, illustrating our survey's capability to detect  $>$0.7\,R$_{\Earth}$ exosatellites.
Our findings reveal no substantial ($>$5$\sigma$) evidence for individual transit events. However, an unusual fraction of light curves favor the transit model at the 2-3$\sigma$ significance level, with fitted transit depths consistent with terrestrial-sized (0.7-1.6\,R$_{\Earth}$) bodies. Comparatively, fewer than 2.2\% of randomly generated normal distributions from an equivalent sample size exhibit a similar prevalence of outliers. Should one or two of these outliers represent a real exosatellite transit, it would imply an occurrence rate of { $\eta = 0.61^{+0.49}_{-0.34}$} short-period terrestrial exosatellites per system, consistent with the known occurrences rates for both solar system moons and mid M-dwarf exoplanets. 
We explore alternative astrophysical interpretations for these outliers, underscoring that transits are not the only plausible explanation. 
For orbital periods $<$0.8\,days, the typical duration of the light curves, { we constrain the occurrence rate of sub-Neptunes to $\eta<$0.35 (95\% confidence) and, if none of the detected outlier signals are real, the occurrence rate of terrestrial ($\sim$Earth-sized) exosatellites to $\eta<$0.51 (95\% confidence)}.
Forthcoming JWST observations of substellar light curves will enable detection of sub-Io-sized exosatellites, allowing for much stronger constraints on this exosatellite population.

\end{abstract}


\section{Introduction} \label{sec:intro}

The Solar System is home to more moons than planets, with over 200 moons surrounding the four outer gas giant planets. 
Of these moons, 15 are relatively massive, $\gtrsim$\,$10^{-6}$\,M$_{\rm planet}$. 
Within our own solar system, it is clear that the process of planet formation frequently results in the production of moons, therefore it can be expected that exoplanetary systems will also host a large number of moons \citep{2006Natur.441..834C,2012ApJ...753...60O, Heller_2014, Heller_2016, Miguel_2016, 2018MNRAS.475.1347M, 2018MNRAS.480.4355C, 2020A&A...633A..93R, 2020MNRAS.499.1023I, cilibrasi2020nbody}.

However, exomoon detection and confirmation has been notoriously difficult, and to date, there are no confirmed exomoons. One possible path for exomoon detection is to search for exomoons orbiting directly-imaged exoplanets using the same techniques that astronomers are already are using to search for exoplanets around nearby stars. 
This could include looking for exomoons via direct imaging,
radial-velocity monitoring, astrometric variations,
and transits as explored by earlier authors
\citep{Lazzoni_2020, Vanderburg_2018, Agol_2015, Cabrera_2007, Heller_2016}. But directly-imaged exoplanets are often shrouded in light from their host star, making exomoon detection challenging using these conventional detection methods.
However not all exoplanets orbit their star in close proximity.
Exoplanets in wide orbits (e.g., HD~203030b, Ross 458(AB)c, COCONUTS-2b; \citealt{2006ApJ...651.1166M,2010MNRAS.405.1140G,2021ApJ...916L..11Z}), unbound free-floating planetary mass objects (FFPs) and larger solitary substellar worlds can be observed without the glaring light of a nearby host-star.
In the near-infrared (NIR), young exoplanets and FFPs shine brightly from the heat of their formation. These young substellar worlds supply an opportunity for the discovery of exomoons orbiting wide-orbit exoplanets or moon-like objects orbiting FFPs and low-mass brown dwarfs (which we refer to as \textit{exosatellites}).

Surveys in the $\lambda > 1 \, \mathrm{\mu m}$ range can more easily detect the exosatellites of NIR-bright hosts.
Several upcoming and ongoing efforts are underway to survey substellar light curves for transiting exosatellites. 
The SPECULOOS survey, which operates just short of $\lambda = 1 \, \mathrm{\mu m}$ can detect terrestrial exoplanets around nearby ultracool stars — of spectral type M7 and later, including some brown dwarfs \citep{2018SPIE10700E..1ID,2020MNRAS.495.2446M}. 
Shifting to the NIR, redward of $\lambda = 1 \, \mathrm{\mu m}$, PINES is an ongoing survey for transiting exosatellites around a sample of almost 400 nearby brown dwarfs and FFPs \cite{2019PASP..131k4401T}. PINES is sensitive to the detection of $>$2.5\,R$_\earth$ exosatellites and has identified one $\sim5$\,$R_\earth$ candidate orbiting 2MASS J0835+1953 \citep{2022AJ....163..253T,2022AJ....164..252T}.
Future transiting exosatellite surveys hold promise for detecting smaller radii companions. It is estimated that the Nancy Grace Roman Space Telescope Galactic Bulge Time Domain Survey may identify a few transiting exosatellites around brown dwarfs \citep{2023arXiv230309959T}. Moreover, the proposed TEMPO survey of the Orion Nebular Cluster with the Roman telescope \citep{2023PASP..135a4401L} would be able to detect dozens of exosatellites around young brown dwarfs and FFPs with sensitivity down to Titan-sized worlds. More imminently, ongoing JWST spectrophotometric observations of FFPs and brown dwarfs could reveal transiting exosatellites down to the size of Jupiter's moon Io \citep{Limbach2021}.

As it happens, there exists another abundant source of substellar light curves with sensitivity sufficient to detect exosatellites, and with a typical observation per target nearing a full day in duration. During its mission, the {\it Spitzer Space Telescope} (hereafter Spitzer) monitored the light curves of dozens of substellar objects, FFPs and wide-separation exoplanets, in most cases with the goal of understanding weather on these worlds. The ability of Spitzer to detect exosatellites transiting substellar worlds has long been known \citep{2013arXiv1304.7248T}. \cite{2017MNRAS.464.2687H} searched for transits in the Spitzer substellar light curves sample from \cite{Metchev2015} using a sliding box technique and under the assumption that all hosts were 60\,M$_{\rm Jup}$, 0.9\,R$_{\rm Jup}$ brown dwarfs. We preformed a previous study that explored this possibility using a more complex Gaussian Process + transit model to simultaneously fit transits and host variability in NIR substellar and FFP light curves. In that study, we identified a possible transit of a habitable-zone 1.7\,R$_\earth$ world in the 2MASS~J1119-1137~AB Spitzer light curve, demonstrating that detection of exosatellites transiting FFPs in archival Spitzer light curves may be possible \citep{Limbach2021}. In this manuscript, we expand our search to 32 FFPs and low-mass brown dwarfs ranging in mass from 3-30\,M$_{\rm Jup}$ in search of transiting exosatellite signals, and with an eye towards constraining the occurrence rates of this elusive low-mass host population.

\section{The Sample} \label{sec:methods}

Table \ref{tab:TargetList} gives the characteristics of the sample examined in this study. It consists of 44 archival Spitzer light curves of L and T dwarfs, with a median estimated host mass of 16\,M$_{\rm Jup}$, inclusive of 14 free-floating planetary-mass objects (FFPs) and one wide-orbit exoplanet (Ross 458c). Our sample includes the targets from \cite{Schneider_2018,Vos2018,Vos2020,Vos2022}, and a reanalysis of the low-mass brown dwarfs and FFPs in the \cite{Metchev2015} sample that were assumed to be 60\,M$_{\rm Jup}$ in a previous transit search (see \citealt{2017MNRAS.464.2687H}). The masses of our targets were sourced from the literature (refer to Table \ref{tab:TargetList} for specific references) which typically derived masses from a combination of moving group associations, parallaxes, spectroscopic data, and photometric observations for each object. In this study, our focus is solely on the lowest mass hosts. Consequently, we carefully included only those hosts with archival Spitzer light curves whose masses, from the literature, are $<$30M$_{\rm Jup}$.  

{ The light curves utilized in this study were initially observed to investigate variability. Although many of these light curves were selected due to their youth or low mass, none of the targets were suspected to host transiting exosatellites prior to the Spitzer observations. Therefore, this sample is unbiased in this regard.}

Whilst a few of our targets were solely observed in the Spitzer [3.6] band, a significant proportion of them were observed in both the Spitzer [3.6] and [4.5] micron bands, as listed in Table \ref{tab:TargetList}. The median light curve observations on each target was 21\,hours. We conducted transit searches in the light curves from each spectral band separately. Consequently, our sample comprises a total of 44 light curves for the 32 targets.

\startlongtable
\begin{deluxetable*}{llllllllcl}
\tabletypesize{\scriptsize}
\tablecolumns{10}
\tablewidth{0pt}
\setlength{\tabcolsep}{0.05in}
\tablecaption{Sample Parameters}
\tablehead{  
\colhead{Name} &
\colhead{SpT} &
\colhead{[3.6]} &
\colhead{Distance} &
\colhead{Age} &
\colhead{$T_\mathrm{eff}$} &
\colhead{Radius} & 
\colhead{Mass} &
\colhead{LC Dura.$^a$} &
\colhead{Ref}\\
\colhead{} &
\colhead{} &
\colhead{(mag)} &
\colhead{(pc)} &
\colhead{(Myr)} &
\colhead{(K)} &
\colhead{($R_{\mathrm{Jup}}$)} & 
\colhead{($M_{\mathrm{Jup}}$)} &
\colhead{(hr)} &
\colhead{} 
}
\startdata
2MASS J0030300-145033   & L7 & 13.37 &  24.3$\pm$2.8    & 30--50           & $1264\pm78$  & $1.36\pm0.04$      & $10\pm2$  & 21/0  & 1  \\
2MASS J00310928+5749364 & L9 & 11.96 &  14.0$\pm$0.9   & 150--250         & $1343\pm58$  & $1.2\pm0.06$       & $24\pm7$ & 21/0  & 1 \\
2MASS J00452143+1634446 & L2$\gamma$ & 10.8 & 17.5$\pm$0.6 & 50 &1970 ± 70 & 1.62$\pm$0.06 &25.0±4.6& 7/7 &  2, 3, 4\\
2MASSI J0103320+193536 & L6 & 12.93& 21.3$\pm$3.0 & 300$\pm$200& 1231 ± 117 & 1.34$\pm$0.13&  15 ± 11 & 14/7 & 2, 5, 6\\
2MASS J01531463-6744181  & L2 & 13.35 & 50.3$\pm$2.9  & 41--49          & $1664\pm53$  & $1.48\pm0.03$      & $17\pm2$    & 20/0  & 1  \\
2MASS J03264225-2102057 & L5$\beta$ & 12.56 & 23.4$\pm$1.1  & 110--150        & $1359\pm46$  & $1.32\pm0.06$      & $18\pm4$  & 21/0   & 1 \\       
2MASS J0342162-681732  & L4$\gamma$ & 13.58 & 51.5$\pm$2.8  & 41--49            &$1650\pm51$  &$1.48\pm0.03$       & $17\pm2$  &  20/0  & 1\\
2MASS J03552337+1133437 & L5$\gamma$ & 10.34 & 9.1$\pm$0.1   & 110--150      & $1527\pm14$  & $1.22\pm0.02$      & $28\pm2$  &  17/0  & 1  \\
2MASS J04590034-2853396 & L7: & 13.92 & 31.1$\pm$2.7   &30--50            & $1152\pm98$ & $1.32\pm0.03$      & $8\pm1$  & 22/0  & 1 \\
2MASS J05012406-0010452 & L3$\gamma$ & 11.77 & 19.6$\pm$1.3 & 300$\pm$200 &1720$\pm$55&  1.38$\pm$0.18 &29$\pm$16&  20/20 & 2, 3\\
2MASS J05065012+5236338 & T4.5 & 14.25 & 16.8$\pm$1.4  & 500-1000                & $1065\pm63$  & $1.05\pm0.08$       & $26\pm7$  &  22/0  & 1  \\
2MASS J06420559+4101599 & L9(red) & 12.89 &  16.0$\pm$0.8  & 110--150        & $1124\pm11$  & $1.23\pm0.0$       & $12\pm1$  &  21/0 & 1  \\
2MASS J07180871-6415310 & T5 & 15.44 & 9.3$\pm$0.9    & 16--28         & $596\pm32$   & $1.29\pm0.02$      & $3\pm1$  & 18/0  & 1 \\
2MASS J08095903+4434216 & L6p & 13.02 & 23.6$\pm$1.8     & 10--150        & $1253\pm74$ & $1.35\pm0.11$      & $13\pm8$    & 21/0  & 1\\
2MASS J09512690 -8023553 & L5(pec) & 14.55 & 58.1$\pm$4.4   & 30--50    &  $1463\pm65$ & $1.45\pm0.05$       & $ 15\pm2$  &  20/0  & 1  \\
2MASS J11193254-1137466AB & L7$\gamma$+L7$\gamma$ & 13.5 & 26.4$\pm$7.1 & 10 ± 3 & 1290$\pm$40 &1.38$\pm$0.14&3.7$\pm$1.1 &  10/10 & 7, 8\\
WISEA J114724.10-204021.3 & L7$\gamma$ & 13.7 & 31.3 ± 3.8& 10 ± 3& 1150$\pm$50 &  [1.4]$^b$  & 5.5$\pm$0.5  & 10/10 &7, 9\\
Ross 458c & T8 & 15.43 &11.7$\pm$0.2&125$\pm$75& 649 ± 14 & 1.142 ± 0.004 & 6.8 ± 0.3 & 14/7 & 2, 5, 6, 10\\
2MASS J13243559+6358284 & T2.5$^*$ & 12.63 &  21.8$\pm$1.7& 130$\pm$20&1080 ± 60&1.23 ± 0.02&11.5$\pm$0.5& 14/7 & 2, 5, 11, 12\\
2MASS J14252798-3650229 & L4$\gamma$ & 11.0 & 11.6$\pm$0.1&130$\pm$20 & 1535 ± 55  & 1.32$\pm$0.09 & 20.8 ± 7.9 &  5/6 &  2, 6\\
SDSS J151643.01+305344.4 & T0.5 & 13.68 & 31 ± 6 & 40-1000 & 1050$\pm$50 & 1.2 & 20$\pm$10& 14/7 & 2, 5, 13\\
2MASS J15515237+0941148  & L4$\gamma$ & 13.24 & 45.2$\pm$2.9   & 10--150        & $1653\pm126$ & $1.39\pm0.19$      & $24\pm14$   & 20/0 & 1 \\
2MASS J16471580+5632057 & L7 & 13.18 & 23.4$\pm$1.1  & 150--250           & $1244\pm47$  & $1.22\pm0.07$      & $18\pm6$   &  20/0  & 1  \\
2MASSI J1726000+153819 & L3$\beta$ & 12.81 & 35$\pm$3.2&300$\pm$200 & 1840$\pm$65& 1.40$\pm$0.20 &28$\pm$17& 14/7 & 2, 3, 5 \\
2MASS J17410280-4642218 & L6-L8$\gamma$ & 11.94 & 19.8$\pm$1.1  & 110--150       & $1471\pm44$  & $1.23\pm0.02$      & $26\pm3$  &  17/0  & 1  \\
2MASS J20025073-0521524  & L5$\beta$ & 12.16 & 17.6$\pm$0.5    & 10--150       & $1301\pm56$  & $1.36\pm0.11$      & $13\pm8$  &  20/0  & 1  \\
2MASS J21171431-2940034  & T0 & 13.01 & 19.1$\pm$2.2 & 16--28   & $1192\pm80$  & $1.38\pm0.05$      & $7\pm2$& 21/0 &   1 \\
HN PegB & T2.5 & 13.72 & 18.4$\pm$0.3& 237$\pm$33 &  1043 ± 23 &1.17 ± 0.05 &20.2 ± 9.4& 14/7 & 2, 5, 6\\
2MASS J22062520+3301144 & T1.5 & 14.63 & 27.9$\pm$1.8& 30--50          & $1076\pm37$  & $1.3\pm0.02$       & $8\pm1$    &  20/0 & 1    \\
2MASSW J2208136+292121 & L3$\gamma$ & 13.26 & 47.2$\pm$1.5&23$\pm$3 &1840$\pm$65&1.41$\pm$0.20& 13$\pm$2& 14/7 & 2, 3, 5 \\
WISE J221628.62+195248.1 & T3  & 14.18 &  22.6$\pm$1.8 &   16--28      & $909\pm41$   & $1.32\pm0.02$      & $5\pm1$ & 22/0 &   1   \\
2MASS J23225299-6151275  & L2$\gamma$ & 12.89 & 43.1$\pm$1.8  & 41--49       & $1740\pm43$  & $1.5\pm0.03$       & $18\pm2$  & 21/0 &   1   
\enddata
\label{tab:TargetList}
$^{a}$ Some observations included light curves in two Spitzer bands. The duration of the light curves is listed in both bands as [3.6] band observation duration / [4.5] band observation duration.
$^{b}$ The radius estimate for WISEA J114724.10-204021.3 was not available in the literature. We estimate the FFP's radius from \cite{2007ApJ...659.1661F} models. 
{\it References:} (1) \citealt{Vos2022} {\it and references therein}, (2) \citealt{Vos2020}, (3) \citealt{Zapatero2014}, (4) \citealt{2019AJ....157..247R}, (5) \citealt{Metchev2015}, (6) \citealt{2015ApJ...810..158F}, (7) \citealt{Schneider_2018}, (8) \citealt{Best_2017}, (9) \citealt{Schneider_2016}, (10) \citealt{Dupuy_2013}, (11) \citealt{2007AJ....134.1162L}, (12) \citealt{Gagn__2018}, (13) \citealt{2009ApJ...702..154S} .
\end{deluxetable*}

\begin{figure}
\centering
\includegraphics[width=0.48\textwidth]{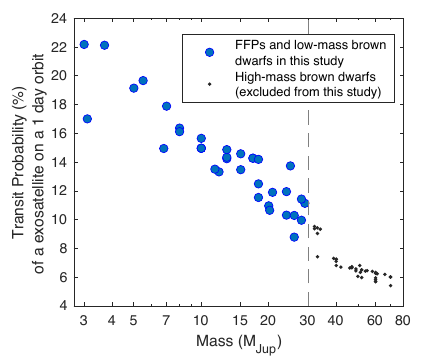}
\caption{Transit probability of an exosatellite on a 1-day orbit as a function of host mass. The colored blue circles denotes the targets in this sample, including all hosts with masses $<$30M$_{\rm Jup}$.  The median transit probability of our sample is 14.3\%, and increases sharply inversely with host mass. Hosts with masses $>$30M$_{\rm Jup}$ (black dots) where excluded from this study -- see \cite{2017MNRAS.464.2687H} for transit search of higher mass ultracool dwarfs.}
\label{Tprob}
\end{figure}

Using the estimated physical parameters of the targets from the literature, we compute the probability of an exosatellite transiting our hosts, $t_{\rm prob}$. The transit probability is given by $t_{\rm prob} \approx R/a$, where $R$ is the radius of the host and $a$ is the semi-major axis of the exosatellite. Using this equation and Kepler's third law gives
\begin{equation}\label{tprobEQ}
    t_{\rm prob} = 0.24R\left(MT^2\right)^{-1/3},
\end{equation}
where $M$ is the mass of the host in Jupiter masses, $R$ is the radius of the host in Jupiter radii and $T$ is the orbital period of the exosatellite in days.

It is noteworthy that our sample includes a confirmed unresolved substellar binary and two more candidate binary systems. For the confirmed binary, we computed transit probability by assuming two objects of equal size and mass in these systems. We treat the binaries as two independent chances to find an exosatellite.

Figure \ref{Tprob} shows the transit probability of an exosatellite on a 1-day orbit assuming random viewing angles { and using the simple approximation $t_{\rm prob} \approx R/a$.  In Section \ref{sec:TransitSearch}, we calculate occurrence rates using the more precise formula $t_{\rm prob} = (R + R_{\rm Sat})/a$. Although our sensitivity varies, we find that $R/a$ (as plotted in Figure \ref{Tprob}) provides a good approximation once completeness factors are accounted for due to our general insensitivity to grazing transits.}
The transit probability experiences a sharp increase for the lowest mass objects. As we have previously indicated \citep{Limbach2021}, this feature is attributable to the young age and low density of these low-mass FFPs, which lead to a markedly higher probability of transits.

We can further calculate the expected transit duration for exosatellites with a given orbital period. We have generated Figure \ref{Tdur}, which illustrates the expected transit duration for and edge-on exosatellite with a period of 0.5 days (gray square) and 3 days (red diamond) for each of our targets. 
Figure \ref{Tdur} reveals distinct patterns: around FFPs, we anticipate longer transit durations, ranging from 1 to 2.5 hours. Mid- to high-mass brown dwarfs are likely to exhibit shorter transit durations, spanning 10 to 60 minutes. The longer transit durations of FFPs and low-mass brown dwarfs produce higher signal-to-noise transits when compared to higher mass brown dwarfs of similar brightness.

The sample consists of young (10-1000\,Myr) L- and T-dwarfs. The targets range in brightness from $10.4-15.4$\,mag in the Spitzer [3.6] band, with the brightest among them being similar in brightness to Trappist-1, which is 10.1\,mag. Despite the lower mass, because of their youth the targets have a range in magnitude similar to more massive field brown dwarfs that have light curves observed with Spitzer. The combination of high transit probability, long transit durations and brightness makes low-mass brown dwarfs and FFPs an ideal substellar population for transit searches.

\section{The Transit Search} \label{sec:TransitSearch}

\subsection{Light curve Construction}\label{Construction}

The light curves used in this study were obtained using the Infrared Camera \citep[IRAC;][]{Fazio2004} on the Spitzer Space Telescope and previously published in the literature, most notably the large surveys published by \citet{Metchev2015} and \citet{Vos2022}. In Table \ref{tab:TargetList} we list the Spitzer light curve reference for each target. 
Despite coming from a range of programs, the targets in this study were observed using standard observing practices. Science images were obtained in staring mode for all targets, while exposure times and observing durations differ between targets. Data were reduced in a similar way across all of the programs: photometry was measured from the Basic Calibrated Data images produced by the Spitzer Science Center. Aperture photometry was performed on targets and reference stars using a variety of sizes, the final chosen aperture producing the lowest rms light curve. 

Raw Spitzer/IRAC time-series photometry is dominated by well-characterized intra-pixel sensitivity variations, whereby a source appears brighter if it is positioned near the center of an IRAC pixel compared to the outer pixel boundaries. When coupled with pointing drifts, the intra-pixel sensitivity variations can induce significant photometric variability that is correlated with $x$ and $y$ sub-pixel coordinates. The light curves used in our study have been corrected for this effect using a quadratic \citep[e.g.][]{Metchev2015} or cubic \citep[e.g.][]{Vos2022} function of the $x$ and $y$ sub-pixel coordinates or by using pixel-level decorrelation (PLD) method developed by \citet{Deming2015, Benneke2017} \citep[e.g.][]{Schneider_2018}.

\begin{figure}
\centering
\includegraphics[width=0.463\textwidth]{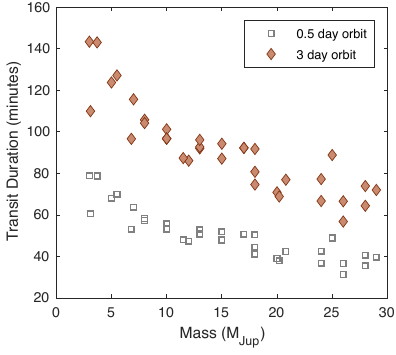}
\caption{Edge-on transit durations for an exosatellite on a half day (gray open squares) and 3\,day (red solid diamonds) orbit about each of FFP and brown dwarf in this sample. Typical transit durations range between 30\,min -- 2.5\,hr for short orbit edge-on exosatellites transiting FFPs and low-mass brown dwarfs.}
\label{Tdur}
\end{figure}

\subsection{Light curve Fitting Methods} \label{sec:LCmethods}

The central goal of this study is to utilize archival Spitzer light curves to establish constraints on the occurrence rate of exosatellites that transit low-mass substellar worlds. To achieve this objective, we will apply two models to all the light curves: one that includes a transit and one that does not. For both models, we use a Gaussian process (GP) model to account for host variability. Subsequently, we calculate the Bayesian evidence for both model fits to determine the $\Delta {\rm log}z$, given by
\begin{equation}
    \Delta {\rm log}z = {\rm log}z\left({\rm GP+transit}\right)-{\rm log}z\left({\rm GP~only}\right),
\end{equation}
which  enables us to ascertain whether any light curves demonstrate a preference for the GP+transit model rather than the GP-only model, which may indicate the potential detection of a transit. 

To fit the transit, we employ a trapezoidal transit model without limb darkening, as the limb darkening parameters are not favored due to the low signal-to-noise of our light curves. The transit model thus uses four parameters: mid-transit time, transit depth, transit duration and the impact parameter. For the majority of impact factors, we expect transit durations between 20\,min--2.5\,hr (see Figure \ref{Tdur}), and use this range as the bound for transit duration.

Our GP model used to fit the substellar variability is akin to that described by \citet{2018MNRAS.478.4866V}. We employ a quasi-periodic kernel:
\begin{equation}
\label{covar}
C_{ij} = h^2 \exp{\left[-\frac{(t_i-t_j)^2}{2\tau^2}-\Gamma \sin^2{\frac{\pi(t_i-t_j)}{T}}\right]}+\left[\sigma_i^2+\sigma_{\text{jit}}^2\right]\delta_{ij}
\end{equation}
where $C_{ij}$ is the covariance matrix; $\delta_{ij}$ is the Kronecker delta function;
$h$ is the amplitude of correlated noise; $t_i$ is the time of $i$th observation; $\tau$ is the correlation periodicity; $\Gamma$ is the ratio between the squared exponential and periodic parts of the kernel; $T$ is the period of the correlation; and
$\sigma_{\text{jit}}$ is a white noise term in addition to the reported uncertainty $\sigma_i$. 
We imposed Jeffreys priors, i.e., log-uniform distributions
for all parameters.
We adopted the following likelihood function:
\begin{equation}
\label{likelihood}
\log{\mathcal{L}} =  -\frac{N}{2}\log{2\pi}-\frac{1}{2}\log{|\bf{C}|}-\frac{1}{2}\bf{r}^{\text{T}}\bf{C} ^{-\text{1}} \bf{r}
\end{equation}
where $N$ is total number of measurements; $\bf{C}$ is the covariance matrix defined earlier; and $\bf{r}$ is the residual of observed flux minus the trapezoid transit model.

Prior to fitting our models, we employed a binning technique to enhance computational efficiency. For light curves spanning over 15\,hrs, we utilized 120 bins, while for those under 15\,hrs, we used 100 bins, thereby ensuring a minimum temporal resolution of 10\,min in all cases. We conducted extensive testing, analyzing multiple light curves at various temporal resolutions, including a higher number of bins. Through this process, we confirmed that the binning procedure did not influence the $\Delta {\rm log}z$ results obtained.

We computed and compared the Bayesian evidence of both models (GP-only and GP+transit) using \texttt{Dynesty}, a nested sampling code \citep{2020MNRAS.493.3132S}. 
For the GP-only model, we utilized 60 live points and specified a \texttt{dlogz} tolerance of 0.2, while keeping the remaining variables at default settings. To calculate the ${\rm log}z$, we determined the average ${\rm log}z$ from 10 runs. In the case of the GP+transit model, which had 10 variables instead of 6, we employed 4000 live points and specified a \texttt{dlogz} tolerance of 0.1. The rest of the variables were set to their default settings. To calculate the ${\rm log}z$, we took the average ${\rm log}z$ from 3 runs. The \texttt{Dynesty} sampling procedure terminated automatically after the \texttt{dlogz} tolerance convergence criteria were met. We selected this approach as it maximized the efficiency of the code and provided a similar level of precision in the final ${\rm log}z$, with an uncertainty of $\pm$0.1, for both models. We conducted a thorough examination and determined that modifying these parameters in \texttt{Dynesty} did not significantly alter the resulting ${\rm log}z$. The Bayesian evidence from the two model fits was used to compute the $\Delta {\rm log}z$.

\begin{figure*}
\centering
\includegraphics[width=0.94\textwidth]{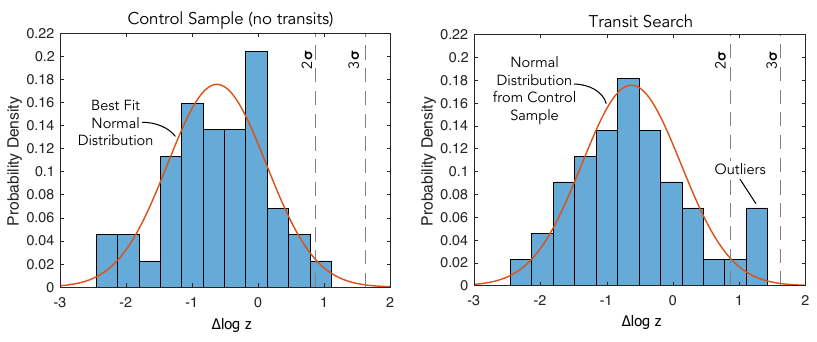}
\caption{Computed probability distributions of $\Delta {\rm log}z$. Light curves with a positive $\Delta {\rm log}z$ favor the GP+transit model, whereas a negative $\Delta {\rm log}z$ is indicates a light curve that favors the GP-only model. {\bf Left:} Distribution for flipped light curves, which contain no transits and is used as the control sample to define a normal distribution. Significant (2-3\,$\sigma$) outliers are defined based on the normal distribution fit to the control sample.{\bf Right:} Distribution for light curves (with no flip). This $\Delta {\rm log}z$ distribution contains several outliers favoring the GP+transit model at the 2-3\,$\sigma$ level not present in the control sample.}
\label{logzDist}
\end{figure*}

\subsection{Results of Transit Search} \label{sec:resultsDetect}

Detecting transits in substellar light curves presents a challenge as they are often affected by host variability caused by inhomogeneous cloud cover and evolving ``weather'' \citep[e.g.][]{Metchev2015,Vos2022}. The GP model attempts to capture the host variability, but differentiating between host variability and a transit event is still a challenging task. {\it To what extent can host variability mimic transits? Can variability alone lead to a preference for our GP+transit model? }
Typically, the Bayes Factor might help differentiate between the GP+transit and GP-only models at certain evidence levels. However, host variability itself could create transit-like fading events, potentially misleadingly favoring the GP+transit model. Therefore, identifying a specific $\Delta {\rm log}z$ threshold that signifies a transit detection beyond mere host variability imitation is crucial.

To assess how variability might imitate transits necessitates a control sample of light curves from substellar objects devoid of transit events. To obtain this control sample, we inverted Spitzer light curves, effectively flipping any transits, ensuring detections in this control sample are attributable to host variability, as our transit model detects dimming, not brightening. This method helps quantify the potential for host variability to mimic transits. We then analyzed the control sample's $\Delta {\rm log}z$ distribution, fitting a Normal distribution to understand the typical behavior in the absence of real transits.

The left panel of Figure \ref{logzDist} shows the distribution of $\Delta {\rm log}z$ values from inverted light curves, with a Normal fit (red line) indicating a mean of -0.63 (where negative values favor the GP-only model) and a standard deviation of 0.75. Based on this control sample, we define 2$\sigma$ and 3$\sigma$ transit detection thresholds for $\Delta {\rm log}z$ at 0.87 and 1.62, respectively. Notably, in the control sample, no light curves exhibited a preference for the transit model with $>$2$\sigma$ significance.

Having established the transit detection threshold from the control distribution, we proceeded to search transit events within the light curves. We applied GP-only and GP+transit fits to the original, unflipped light curves, calculating the $\Delta {\rm log}z$ values. The distribution of these values, depicted in the right panel of Figure \ref{logzDist}, deviates from the control, notably due to an increase in outliers favoring the GP+transit model. Among the 44 light curves, we identified three outliers with $>$2.4$\sigma$ significance. We find that for randomly generated Normal distributions of the same sample size, this occurs less than 2.2\% of the time. Thus, there are three possible explanations (or some combination of these explanations) that can explain the observed outliers in the light curves:
\begin{enumerate}
    \item Random chance (2.2\%)
    \item The presence of an exosatellite transit(s) within the sample that are marginally detectable
    \item Different host variability behavior between brightening and dimming events (see discussion below)
\end{enumerate}
When using the flipped light curves as a control sample, the underlining assumption is that host variability produces brightening and fading events that are similar. If, however, host variability produces fading events that inherently resemble transits more so than (flipped) brightening events do, then our control sample might not provide an accurate transit detection threshold. This offers an alternative astrophysical explanation for the observed $\Delta {\rm log}z$ outliers in the distribution without necessitating actual transits. 

\begin{figure*}
\centering
\includegraphics[width=0.99\textwidth]{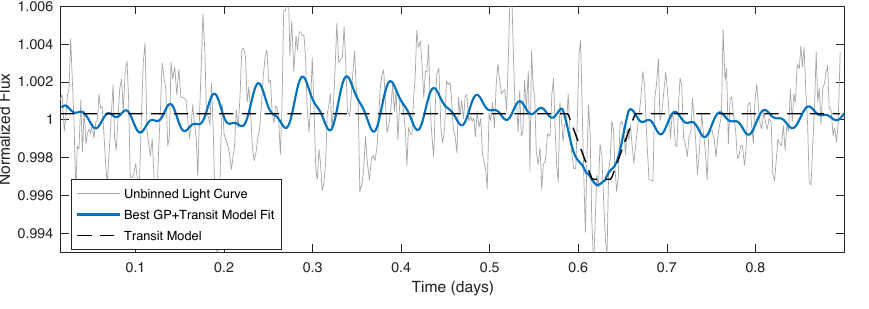}
\caption{The Spitzer 3.6\,$\mu$m  light curve for 2MASS J21171431-2940034, spanning a duration of 21 hours, illustrated by the light gray line.  Overlaid is the best fit GP+transit model (shown in blue), as well as the isolated transit model (displayed as a black dashed line). The 2.7$\sigma$ transit event occurs just past the 0.6-day mark. The transit model that best fits this event has a transit depth of 0.34$\pm$0.12\% or 0.77$\pm$0.26\,R$_\Earth$. Using the best fit impact parameter, transit duration (65$\pm$25\,min) and host mass (7$\pm$2\,M$_{\rm Jup}$), we derive an orbital period of 1.5$^{+2}_{-1}$\,days, similar to Io and Trappist-1b (although there is very significant uncertainly).}
\label{2m2117LC}
\end{figure*}

\begin{figure}[b]
\centering
\includegraphics[width=0.5\textwidth]{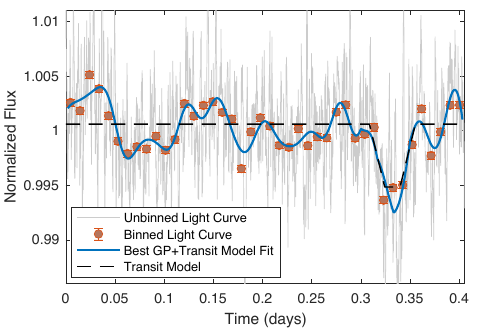}
\caption{Displayed is the Spitzer 3.6\,$\mu$m light curve for 2MASS J11193254-1137466AB over a span of 10 hours, depicted by the light gray line. The data is also binned for clarity, and the corresponding binned data points are marked by red error bars. Superimposed on the plot is the GP+transit model that best fits the data (shown in blue), along with the isolated transit component (depicted by the black dashed line). The transit event with a significance of 2.6$\sigma$ is observed between the time marks of 0.3 and 0.35 days. The transit fit has a depth of 0.59\% and a duration of 43 minutes. This event is described in more detail in \cite{Limbach2021}, the only difference here being the inclusion of the impact parameter in our transit model which changed the best fit parameters slightly.}
\label{2m1119LC}
\end{figure}

{ One or more transiting exosatellites} near the detection threshold may also explain the observed distribution.
The three targets that most significantly favor the transit model are 2MASS J21171431-2940034 (2.7$\sigma$), 2MASS J11193254-1137466AB (2.6$\sigma$; previously published by \citealt{Limbach2021}) and 2MASS J06420559+4101599 (2.4$\sigma$). The two light curves that favor the GP+transit model at $>2.5\sigma$ are shown in Figures \ref{2m2117LC} and \ref{2m1119LC}. 

Notably, all three targets are $\leq$12\,M$_{\rm Jup}$, even though only 44\% of our sample is planetary-mass. A plausible interpretation supporting the presence of transits is that FFPs with lower masses exhibit high transit probabilities (see Figure \ref{Tprob}), resulting in transits that both occur more often and that are easier to detect. Nonetheless, an alternative explanation could be that these FFPs are young and thus display enhanced or irregular variability \citep{Vos2022}. The heightened variability of young brown dwarfs may potentially contribute to a greater occurrence of false positives in this sample. However, if this is the case, one might also expect this to also be present in the flipped light curves of FFPs, which we do not observe.

It is observed that GP models have difficulty distinguishing between two types of events when their frequencies are similar. For 2M1119 and 2M0642, the fitted duration of the transit is similar to the rotation period of the host, whereas this is not the case for the 2M2117 light curve. This might imply a higher likelihood that the event observed in 2M2117 is an actual exosatellite transit.

We also searched for larger transiting exosatellites (e.g., sub-Neptunes), with transit depths up to 3\%. Our investigation yielded no statistically significant indications of any exosatellites exhibiting transit depths exceeding 1\%. In a recent study, the PINES survey suggested a potential rise in the occurrence rate of sub-Neptunes around L- and T-dwarfs relative to M-dwarfs \citep{2022AJ....164..252T}. It is worth highlighting that our findings do not contradict their results, as the majority of the PINES survey data pertains to massive brown dwarfs. In contrast, our occurrence rates pertain exclusively to lower-mass brown dwarfs and FFPs. Indeed, we would expect lower-mass brown dwarfs and FFPs to form lower-mass exosatellites \cite{cilibrasi2020nbody}. More specifically, the PINES sub-Neptune transiting candidate is in orbit around a substellar host likely exceeding 60 M$_{\rm Jup}$, more than twice the mass of the most massive brown dwarf in our sample.

\section{Occurrence Rates} \label{sec:OccRates}

\subsection{Injection/Recovery Test} \label{sec:InjectionRecovery}

Even with flipped light curves as a control, discerning whether transit-like detections stem from actual transits or host variability is challenging. While a definitive distinction is difficult, we can ascertain whether (1) we can accurately recover simulated transits resembling those we might have detected, and (2) if the $\Delta {\rm log}z$ values of these recovered transits align with the $\Delta {\rm log}z$ values of similar potential transits in the actual data, using an injection and retrieval test.

First, we examine our ability to recover exosatellite transits with parameters similar to the transits detected with $>$2.4\,$\sigma$ significance. The outlier transits were found to have depths between 0.34-0.9\% and durations of 43-90 minutes. To assess our retrieval accuracy for similar transits, { we injected 426 transits into all flipped light curves,  with durations ranging from 10-150\,min and depths ranging from 0.3-1.0\%, and using a distribution of transit mid-transit times and impact parameters}. We then attempted to recover these transits. 
We consider a retrieval successful if the recovered mid-transit time is within $\pm$15 minutes of the injected time { and the transit detection is $>2\sigma$ significance}.

\begin{figure}[]
\centering
\includegraphics[width=0.5\textwidth]{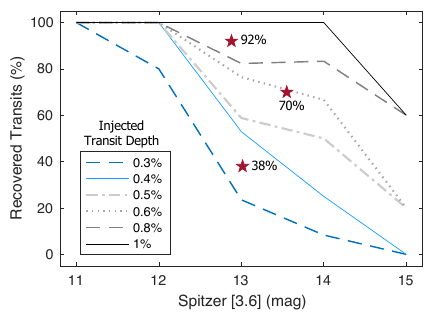}
\caption{Percentage of retrieved transits as a function of the target's Spitzer [3.6] band magnitude for injected transits with transit depths ranging between 0.3-1.0\% { and with a duration of 60\,min}. The red stars correspond to the transit fits that had $\Delta {\rm log}z$ values of $\geq$2.4$\sigma$ significance. The percentage of transits with similar transit depths that are successfully retrieved is labeled on the plot. In all three cases, transits with similar proprieties were retrievable in a substantial (38-92\%) portion of the sample, indicating the data enables detection of transits with parameters comparable to those detected. }
\label{PercentRetriv}
\end{figure}

Figure \ref{PercentRetriv} illustrates the retrieval rate of transits across various target magnitudes { for a transit duration of 60\,min}. For the brightest targets, we successfully recover all injected transits, even those as shallow as 0.3\%. However, retrieval efficiency declines for fainter targets, particularly at minimal transit depths. The plot also features three red stars representing the transits identified as $>$2.4$\sigma$ outliers. In these instances, 38-92\% of transits with comparable injection parameters were accurately retrieved from light curves of similarly bright substellar hosts. Although this does not demonstrate if the detected transits are real or not, it does suggest that transits with similar features are indeed detectable in our sample.

{ Figure \ref{TranDurPlt} shows the recovery rate of transits with different orbital periods. We find we are nearly uniformly sensitive to all orbital periods of interest for the sample as a whole (gray dashed line). This outcome may seem unexpected for those accustomed to identifying transits in Kepler/TESS light curves. The unusual nature of this result can be attributed to two factors: a) the existence of correlated noise in the light curves due to host variability on timescales similar to the transit durations, and therefore b) our very different detection algorithm. More specifically, for the brighter targets, our sensitivity tends to decrease as the transit duration (or orbital period) increases. This pattern suggests that detection of transits in brighter targets is limited more by substellar variability than by photon noise, under the assumption that substellar variability more effectively mimics longer (1-2\,hr or more) transit durations compared to shorter ($\lesssim$30\,min) ones. However, the opposite appears to be true for the fainter targets. Here transit detection appears to be in photon noise limited regime, so increasing the transit durations improves transit detection, which is generally true for exoplanet transits around stars. 

\begin{figure}[]
\centering
\includegraphics[width=0.44\textwidth]{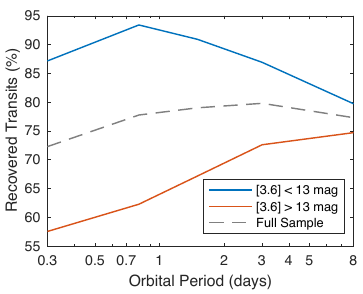}
\caption{{ Percentage of recovered transits with a depth of 0.7\% as a function of orbital period for bright targets (blue line) and fainter targets (red line) and the sample as a whole (gray dashed line).}}
\label{TranDurPlt}
\end{figure}

We recognize that typically, transits would be injected based on specific orbital and physical parameters such as periods, inclinations, and companion radii. However, we deliberately opted for a predefined range of transit depths and durations rather than the more conventional parameters. This decision stems from the fact that the masses or radii of the hosts in our study are often very poorly constrained, which differs significantly from when we study transiting exoplanets around stars. In some cases, the hosts have 1$\sigma$ mass errors of $\sim$400\%. Consequently, using mass for the conversion between transit duration and orbital period/inclination is likely to introduce significant inaccuracies.

Despite the large error bars associated with individual masses in certain cases, the mass distribution of the sample is likely to be more accurate overall, even if the mass of any single object is not reliable. Thus, by seeking transits of the same duration in every light curve, we mitigate the impact of individual mass errors on each target, making us less sensitive to these uncertainties. When we do convert injected transit durations to a given inclination and orbital period, we used the median mass and radii of the sample rather than the individual values. 
As future studies yield better constraints on the mass and radii of substellar worlds and young planets, it would be prudent to reconsider this approach.

Finally, we note that our code currently only has the ability to search for single transits in the lightcurve. When we inject multiple transits for short period satellites, our code detects one of the two transits with similar statistical significance as it detects one transit, and thus multiple transits do not increase (or decrease) detectability. This is an area where future improvements to our code would likely prove beneficial.

}

\begin{figure}[]
\centering
\includegraphics[width=0.39\textwidth]{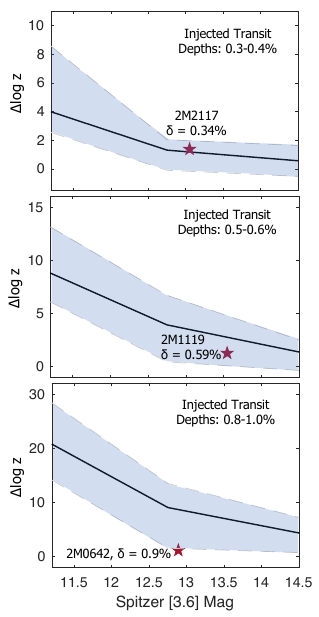}
\caption{Median $\Delta {\rm log}z$ (black line, with blue band showing one standard deviation) as a function of the target's Spitzer [3.6] band magnitude for transits that were successfully injected in retrieved. From top to bottom, the injected transit depths were 0.3-0.4\%, 0.5-0.6\% and 0.8-1.0\%. The red stars (labeled with target name and retrieved transit depth, $\delta$) show the $\Delta {\rm log}z$ of the retrieved transits from the real transit search, all of which lie within one standard deviation of the median retrieved value with the exception of the bottom plot (the 2M0642 retrieval), which lies slightly below the typical values.}
\label{compareLogzs}
\end{figure}

Next, we assess whether the $\Delta {\rm log}z$ values for the detected transits correspond to those retrieved for similarly deep transits around hosts of comparable brightness. Figure \ref{compareLogzs} plots $\Delta {\rm log}z$ against host magnitude for varying transit depths from small (top) to large (bottom). The plots suggest that, possibly with the exception of 2M0642, the retrieved $\Delta {\rm log}z$ values align with those from the injection/retrieval sample. Again, while this doesn't confirm the detected signals as transits, it does indicate that the detected signal is indistinguishable from a transit.

\subsection{Results} \label{sec:ResultsOcc}

We now aim to place some constraints on occurrence rates. The number of exosatellites we expect this survey to detect, denoted as $N_{detections}$, is given by:
\begin{equation}
   N_{detections} = \eta t_{prob}R_c N_{targets},   
\end{equation}
where $\eta$ is the number of exosatellites/target, $t_{prob}$ is the transit probability defined in equation \ref{tprobEQ}, 
$N_{targets}$ represents the total number of substellar bodies observed, and $R_c$, the completion rate is defined as:{
\begin{equation}
  \begin{aligned}
   R_c = \frac{{O_{dur}}}{<T>}\xi \text{    if    }  \frac{O_{dur}}{<T>}<1  \\ 
\text{or   }  
   R_c = \xi \text{    if    }  \frac{O_{dur}}{<T>}\geq1.
     \end{aligned}
\end{equation}
In this equation, $<T>$ represents the average orbital period of interest and $O_{dur}$ is the total observation duration, and $\xi$ is the detection efficiency. The detection efficiency corrects for our sensitivity to specific transit depths and durations. The ratio of the total observation duration to $<T>$ indicates the probability of observing a transit, considering the satellite's orbital period relative to the observation window's length, which is limited to be $\leq$1.} The detection efficiency here is the proportion of detected transits within the transit depth range of interest as determined from our injection/recovery test findings. 

First, we compute the number of exosatellites the survey would have detected if every host had a short-orbit exosatellite (i.e., $\eta = 1$). This provides { an understanding of the sensitivity} of our survey at various transit depths. Using these equations, if every host has an exosatellite with an orbital period of 1\,day, then we would expect to detect the number of exosatellites listed in Table \ref{NoDetections}.
\begin{table}[h]{\label{NoDetections}}
\centering
\begin{tabular}{c|c}
Transit Depth & $N_{expected}$   \\ \hline
0.35\%  & 1.8    \\ 
0.5\% & 2.7 \\
0.6\%   & 3.2     \\ 
0.9\%   &  4.0   \\ 
$>$1\% & 4.4 \\
\end{tabular}
\caption{This table gives the expected number (50th percentile) of exosatellite detections, $N_{expected}$, assuming a scenario where there is 1 exosatellite per substellar world, each with an orbital period of 1 day. The data is organized for various transit depths, which are listed in the left column.}
\label{tab:exosatellites}
\end{table}

\begin{figure*}[]
\centering
\includegraphics[width=0.69\textwidth]{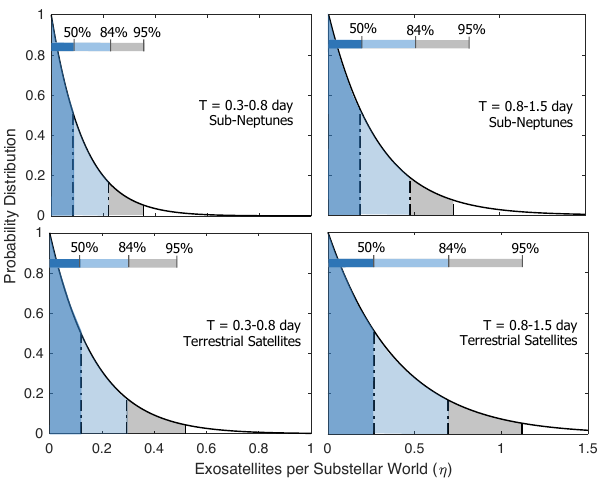}
\caption{Occurrence rate constraints under the assumption that there are no detected transiting exosatellites in this sample. Upper limits on the occurrence rates for sub-Neptunes on 0.3-0.8\,day orbits (upper left panel) and 0.8-1.5\,day orbits (upper right panel). Upper limits on the occurrence rates for terrestrial ($\sim$0.7-1.6\,R$_\Earth$) exosatellites on 0.3-0.8\,day orbits (bottom left panel) and 0.8-1.5\,day orbits (bottom right panel).}
\label{ZeroD}
\end{figure*}

No large ($>$1\%) $\sim$sub-Neptune transits were detected in our survey. 
To determine the occurrence rate of sub-Neptune sized exosatellites based on this survey, we calculate the probability distribution of detecting a certain number of exosatellites per system. We did this by calculating the binomial probability of detecting 0 moons in our survey, assuming an overall average detection efficiency $R_c$ (for such large transits, $R_c \approx 1$) and transit probability $t_{prob}$ for objects in the period/radius range we consider, and many different possible values of the occurrence rate $\eta$.  The calculation of the binomial probability depends on the probability of detecting an exosatellite around any substellar host ($\eta R_c t_{prob}$) and the number of substellar hosts observed. We integrate the probability distribution, applying a uniform prior, and calculate the 1$\sigma$ and 2$\sigma$ limits on the occurrence rate of sub-Neptune-sized exosatellites. 
From the computed probability distribution, shown in the top two panels in Figure \ref{ZeroD}, { we constrain the occurrence rate of $>1\%$ deep ($\sim$sub-Neptune sized) transits to $\eta<$0.35 (95\% confidence) for orbital periods of 0.3-0.8\,days and to $\eta<$0.48 (84\% confidence) for orbital periods of 0.8-1.5\,days. }

We can perform a similar analysis for smaller transit depth signals that correspond to terrestrial ($\sim$0.7-1.6\,R$_\Earth$) companions. We compute occurrence rates for terrestrial exosatellies for two different scenarios: (1) the case where one or two of the outlier transit signals is real, and (2) the case where none of the detected outlier signals correspond to real transits. For case 2, we estimate the occurrence rate by computing the probability distribution using the same method as we did for the larger, sub-Neptune sized transits. We use the detection efficiency from Figure \ref{PercentRetriv}, instead of presuming complete retrieval as was done for the larger sub-Neptune transit depths. We conduct the analysis for transit depths of 0.6\%, corresponding to roughly Earth-sized exosatellites. { This constrains the occurrence rate of $0.6\%$ transits to $\eta<$0.51 (95\% confidence) for orbital periods of 0.3-0.8\,days and to $\eta<$0.68 (84\% confidence) for orbital periods of 0.8-1.5\,days, as shown in the bottom two panels in Figure \ref{ZeroD}.}

If instead, we now assume one or two of the $>$2.4$\sigma$ significance outlier detections is a real transit, we can compute the probability distribution under the assumption that one or two exosatellites were detected. To do this, it is first helpful to compute the orbital period for each of the possible detections. 
We can approximate $a$ by first calculating the orbital velocity of the exosatellite, using the best-fit transit duration, impact parameter, and the host's diameter from Table \ref{tab:TargetList}. We then use the host's mass (also from Table \ref{tab:TargetList}) and the relation $a \approx GM/v^2$ to estimate the semi-major axis. We caution that this approximation has significant uncertainties due to errors in the impact parameter fit, host mass and other parameter uncertainties. To calculate the orbital period, we apply Kepler's third law with the computed semi-major axis, leading to orbital periods of 1.5 days (2M2117), 0.7 days (2M1119), and 3.4 days (2M0642). We limit the maximum possible detections to 2 and do not include the possible detection around 2M0642 in our statistical analysis of occurrence rates as the estimated orbital period of this exosatellite is outside the parameter space in consideration (e.g., 0.3-1.5\,days). 

\begin{figure}[b]
\centering
\includegraphics[width=0.45\textwidth]{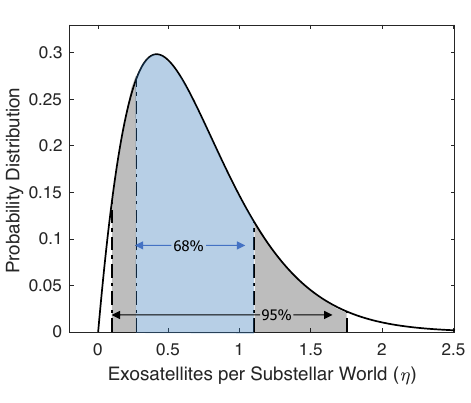}
\caption{Occurrence rate constraints under the assumption that one or two of the detected outlier transits is real. This distribution is the average of the two distributions derived for 1 or 2 outlier detections and constrains the occurrence rate of short-period, terrestrial ($\sim$0.7-1.6\,R$_\Earth$) companions to { $\eta = 0.61^{+0.49}_{-0.34}$} exosatellites per substellar world.}
\label{ProbD1or2}
\end{figure}

We compute the probability distribution of the number of detected exosatellites per substellar world using the the average detection efficiency for any companion in the size range ($\sim$0.7-1.6\,R$_\Earth$) and for orbital periods between 0.3 to 1.5 days. We then calculate the distribution assuming one and then two detections. We then take the average of the two probability distributions (assuming roughly equal probability) which results in the probability distribution shown in Figure \ref{ProbD1or2}. From this distribution, we find that the occurrence rate of short-period, terrestrial ($\sim$0.7-1.6\,R$_\Earth$) companions is { $\eta = 0.61^{+0.49}_{-0.34}$ exosatellites per substellar world } under the assumption that one or two of the outlier events corresponds to a real transit.

To within $\sim$1$\sigma$, this rate aligns with the solar system's moon frequency of 0.25 moons/planet; only Io has a similarly large companion-to-host mass ratio ($5\times10^{-5}$) and short orbital period (1.8\,days). It is also in agreement with the 0.27$^{+0.25}_{-0.18}$ rate for 1.5-2.5\,R$_\Earth$ companions around mid M-dwarfs \citep{2019AJ....158...75H}, which also have similar companions-host mass-ratios and short-orbits. Consequently, if any detected transit is real, especially the 2M2117 or 2M1119 event, then the occurrence rate of exosatellites in this parameter space would be comparable to those of M-dwarf and solar system moon populations. While our findings do not provide definitive constraints, they may hint at potential similarities, with prospects for more precise future determinations from upcoming space-based infrared observatories.

\section{Discussion} \label{sec:discuss}

\begin{figure*}[]
\centering
\includegraphics[width=0.68\textwidth]{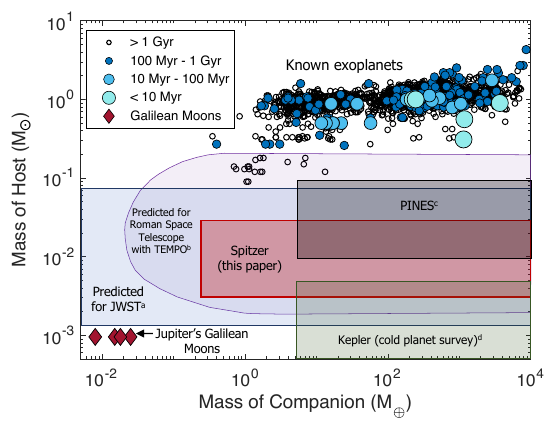}
\caption{Companion vs. host mass for the known exoplanet population (blue and black points) as well as the approximate sensitivity of several past, ongoing or upcoming exosatellite (or moon) surveys of substellar, FFP and exoplanet hosts. This study was sensitive to exosatellites down to the size of 0.7\,R$_\Earth$ ($\sim$0.25\,M$_\Earth$) around 3-30\,M$_{\rm Jup}$ hosts. The majority of the targets within this survey, TEMPO and JWST are young (1-1000\,Myr) unlike the typical stars that host known exoplanets. {\it References:} (a) \citealt{Limbach2021} , (b) \citealt{2023PASP..135a4401L}, (c) \citealt{2022AJ....163..253T,2022AJ....164..252T} and (d) \citealt{2022NatAs...6..367K}.}
\label{paramSpace}
\end{figure*}

To contextualize our exosatellite search, we compared our surveyed parameters with those of other studies. Figure \ref{paramSpace} illustrates the mass of companions versus host mass for known exoplanets and the Galilean moons, alongside the sensitivity range of various past, present, and future surveys targeting exosatellites and/or exomoons. Except for the Kepler survey, which focuses on stellar transits by exomoons, other surveys directly observe the light from brown dwarfs, FFPs, or exoplanets to detect moon/satellite transits. For the Kepler exomoon survey around cold gas giants \citep{2022NatAs...6..367K}, we estimate a minimum detectable exomoon mass of about 5\,M$_\Earth$, derived from a near-threshold sub-Neptune-sized tentative detection. This figure does not encompass the parameter space explored by the earlier Kepler exomoon survey, which could detect far smaller moons but was limited to exoplanets within $<$1\,AU. In this region, theoretical modeling and empirical data suggest a low exomoon formation and occurrence rate \citep{2002ApJ...575.1087B,2010ApJ...719L.145N,2018AJ....155...36T,2020MNRAS.499.1023I}. The PINES and Kepler cold planet surveys demonstrate similar sensitivity to companion sizes; however, the PINES targets are unbound and significantly larger brown dwarf hosts. Operating from the ground, PINES can monitor hundreds of targets \citep{2022AJ....163..253T,2022AJ....164..252T}.

This Spitzer study was sensitive to exosatellites with companion-host mass ratios as low as approximately 5$\times10^{-5}$ — roughly 0.25\,M$_\Earth$ orbiting a median host mass of 16\,M$_{\rm Jup}$ in our sample. In the Solar System, only Io has such a high mass ratio and is on a short (1.8-day) orbit. The upcoming Roman Space Telescope TEMPO survey surpasses Spitzer's capabilities, potentially detecting lower-mass exosatellites around hundreds of young, $<$10 Myr substellar worlds \citep{2023PASP..135a4401L}. Moreover, JWST could identify even smaller moons with mass ratios around 1$\times10^{-6}$ \citep{Limbach2021}, of which there are 15 moons at or above this mass ratio in our solar system. Considering the higher occurrence rate of short-orbit ($<$5 days) moons in the Solar System (1.75 moons/giant planet) and the rate at which JWST is observing these worlds (approximately 5 substellar objects in Cycle 2), the likelihood of detecting an exosatellite or exomoon within the first 5-10 years of JWST operations is promising, especially if their occurrence rate are similar to that of our Solar System moons. The analysis of JWST cycle 1 and 2 light curves for exosatellite transits is already in progress and will be the focus of several forthcoming papers.

\section{Conclusion} \label{sec:conc}

In this study, we searched 44 archival Spitzer light curves of 32 substellar objects with masses $<30~M_{\mathrm{Jup}}$ for transiting exosatellites. It is one of the most sensitive exosatellite surveys to date, capable of discerning transits as small as 0.7\,R$_\Earth$. While no definitive evidence of exosatellite transits was found, an unusually large number of light curves indicated a preference for the GP+transit model at the 2-3\,$\sigma$ level. This could indicate either the presence of terrestrial exosatellite transits hovering near our detection limit or host variability that resembles transit signals. The absence of correspondingly similar transit signals in the flipped light curves favors the transit explanation, provided that host variability doesn't exclusively cause fading events (but not brightening events) similar to transits. Ultimately, it remains challenging to conclusively determine whether transits are present in the observed light curves with current data.

Despite the inconclusive nature of this study's results, the demonstration of Spitzer's ability to detect transiting exosatellites smaller than Earth orbiting FFPs is noteworthy, and very promising for future space-based infrared surveys. Ongoing and forthcoming observations with JWST and the Roman Space Telescope will achieve sensitivity to far smaller exosatellites and exomoons. These missions are likely to detect and confirm the first exosatellites orbiting FFPs and low-mass brown dwarfs, especially if the occurrence rate of these elusive terrestrial worlds resemble those of moons in our Solar System.

\section*{Acknowledgement}
{ The authors thank an anonymous reviewer for a thorough and helpful review of this manuscript.} This research has made use of the NASA Exoplanet Archive, which is operated by the California Institute of Technology, under contract with the National Aeronautics and Space Administration under the Exoplanet Exploration Program. This research has made use of the SIMBAD database, operated at CDS, Strasbourg, France. J. M. V. acknowledges support from a Royal Society - Science Foundation Ireland University Research Fellowship (URF$\backslash$1$\backslash$221932). 

\facilities{{\it The Spitzer Space Telescope}, {\it Gaia}, {\it WISE}, {\it 2MASS}, {\it IRTF/SpeX}}

\software{{\tt SEDkit} \citep{2015ApJ...810..158F}, {\tt astro.py} Astropy { \citep{2013A&A...558A..33A, 2018AJ....156..123A, 2022ApJ...935..167A}}, {\tt numpy.py} \citep{5725236}, {\tt dynesty.py}  \citep{2020MNRAS.493.3132S}, {\tt corner.py} \citep{corner} and {\tt ChatGPT} was utilized to improve wording at the sentence level, answer coding inquires and to convert tables to \LaTeX; Last accessed in January 2024, OpenAI (\url{chat.openai.com/chat}).}

\bibliography{main}{}
\bibliographystyle{aasjournal}
\end{document}